\documentclass[aps,pra,amssymb,amsmath,eqsecnum,nofootinbib]{revtex4}
\usepackage{graphicx}

\newtheorem{definition}{Definition}[section]
\newtheorem{theorem}{Theorem}[section]

\newenvironment{hypothesis}{HP: \begin{center}} {\end{center}}
\newenvironment{thesis}{TH: \begin{center}} {\end{center}}

\newtheorem{example}{Example}[section]
\begin{document}
\title{Some elementary rigorous remark about the replica formalism in the Statistical Physics' approach to threshold phenomena in Computational Complexity Theory}
\author{Gavriel Segre}
\maketitle
\newpage
\section{Introduction}
The adoption of methods and ideas from Statistical Physics in the
analysis of  threshold phenomena in Computational Complexity
Theory \cite{Papadimitriou-94} is a very interesting research
field \cite{Hartmann-Weigt-2005}, \cite{Percus-Istrate-Moore-06}.

Unfortunately some of these methods and ideas has not reached yet
the level of mathematical rigor of Mathematical Physics and
Theoretical Computer Science.

This applies in particular as to the \emph{replica formalism} and
as to the concept of \emph{replica symmetry breaking}.

As prototype  of this situation let us consider the
\emph{satisfability threshold conjecture} for the problem
\emph{random k-SAT}:

given a uniformly-distributed random  boolean formula in
\emph{conjunctive normal form} involving n boolean variables $
x_{1}, \cdots , x_{n}$ and m clauses each of length k (i.e.
containing k literals) such a conjecture states the existence of a
critical value $ \alpha_{c} $ for the \emph{clause density} $
\alpha := \frac{m}{n} = \lim_{n,m \rightarrow \infty} \frac{m}{n}$
such that for every $ \epsilon > 0$  in the limit $ n \rightarrow
\infty $ the probability that a formula is satisfiable tends to 1
if $ \alpha < ( 1- \epsilon) \alpha_{c} $ while tends to 0 if $
\alpha
> ( 1 + \epsilon) \alpha_{c} $.

Introduced the spin variables $ s_{i} := 2 x_{i} -1 $ and
introduced the clause matrix J such that:
\begin{equation}
    J_{ji} \; := \; \left\{%
\begin{array}{ll}
    +1 , & \hbox{if clause j includes the literal $ x_{i}$;} \\
    -1, & \hbox{if clause j includes the literal $ \bar{x_{i}}$;} \\
    0, & \hbox{otherwise.} \\
\end{array}%
\right.
\end{equation}
it follows that that the number of violated clauses may be
expressed as:
\begin{equation}
    H(s_{1}, \cdots , s_{n} ) \; := \; \frac{1}{2^{k}}
    \sum_{j=1}^{\alpha n} \prod_{i=1}^{n} ( 1 - J_{j i} s_{i} )
\end{equation}
The Statistical Physics' approach to the  \emph{random k-SAT}
problem consists in considering H as the hamiltonian of a spins'
system Sys at thermodynamical equilibrium at temperature T whose
canonical partition function (we adopt units in with $
k_{Boltzmann} = 1 $)
 is hence:
\begin{equation}
    Z \; := \; \sum_{s_{1}, \cdots , s_{n}} \exp( - \frac{ H(s_{1}, \cdots , s_{n})}{T}
    )
\end{equation}
Denoting the thermodynamic average over the spins with brackets $
< \cdot > $ and the average over the random instances with a bar $
\bar{\cdot} $ one has then that the averaged number of violated
clauses can be expressed as:
\begin{equation}
    \bar{E} \; :=\overline{ < H > } \; = \lim_{T \rightarrow 0}  \lim_{n \rightarrow \infty} T \overline{\log Z}
\end{equation}
Using the formula:
\begin{equation}
    \log Z \; = \; \lim_{r \rightarrow 0} \frac{Z^{r}-1}{r}
\end{equation}
one can then express $  \bar{E} $ as:
\begin{equation} \label{eq:average number of violated clauses}
     \bar{E} \; = \; \lim_{T \rightarrow 0}
     T  \lim_{n \rightarrow \infty}  \lim_{r \rightarrow 0} \frac{\overline{Z^{r}}-1}{r}
\end{equation}
The first step of the replica formalism consists in expressing $
Z^{r} $, for $ r \in \mathbb{N}_{+}$, as the partition function of
r non-interacting replicas $Sys_{1} , \cdots , Sys_{r} $ of the
system Sys:
\begin{equation}
    Z^{r} \; = \; \sum_{s^{1}_{1} , \cdots s^{1}_{n} } \cdots \sum_{s^{r}_{1} , \cdots s^{r}_{n}
    }  \exp( -  \sum_{a=1}^{r} \frac{ H(s^{a}_{1}, \cdots ,
    s^{a}_{n})}{T} )
\end{equation}
The next step in the replica  formalism consists in "prolonging
analytically" such an expression of $ Z^{r} $ to $ r \in
\mathbb{R}$ and to substitute it into equation\ref{eq:average
number of violated clauses} obtaining:
\begin{equation} \label{eq:nonsense formula number one}
     \bar{E} \; = \; \lim_{T \rightarrow 0}
     T  \lim_{n \rightarrow \infty} \lim_{r \rightarrow 0} \frac{\overline{\sum_{s^{1}_{1} , \cdots s^{1}_{n} } \cdots \sum_{s^{r}_{1} , \cdots s^{r}_{n}
    }  \exp( -  \sum_{a=1}^{r} \frac{ H(s^{a}_{1}, \cdots , s^{a}_{n})}{T})}-1}{r}
\end{equation}

As correctly stated in \cite{Percus-Istrate-Moore-06} as well as
in the section 5.4.3 "The replica approach" of
\cite{Hartmann-Weigt-2005}, anyway, from a mathematical point of
view such a formalism is absolutely nonsense:
\begin{enumerate}
    \item the existence of the thermodynamical limit $ n \rightarrow
    \infty $ is not obvious and has to be proved
    \item assumed the existence of the thermodynamical limit, the fact that for finite n the probability distribution
    of Z is determined by its moments $ \{ \overline{Z^{r}} \, , \, r \in \mathbb{N} \} $  ceases to hold when $ n
    \rightarrow  \infty $
    \item an analytic function is not determined by the values
assumed on a countable set \cite{Hassani-99}.
\end{enumerate}

Hence equation \ref{eq:nonsense formula number one} has no
mathematical meaning.

Actually what one obtains in such a formalism is an expression of
the form:
\begin{equation}
   \overline{ Z^{r} }\; = \; \int \prod_{\sigma} d u_{\sigma} \delta ( \sum_{\vec{\sigma}} u_{\vec{\sigma}} - 1)
    \exp ( n F ( \vec{u} ))
\end{equation}
where $ \vec{\sigma} \in \{ -1 , 1 \}^{r}$ , $ u_{\vec{\sigma}}
\in \mathbb{R}^{2^{r}} $ while $ F :  \mathbb{R}^{2^{r}} \mapsto
\mathbb{R} $ is a suitable function
\cite{Percus-Istrate-Moore-06}.

One then uses a saddle-point approximation of such an integral:
\begin{equation}
     \overline{Z^{r}} \; = \; \exp ( n  \exp F_{max} + o_{n \rightarrow \infty}(n) )
\end{equation}
where with $ o_{n \rightarrow \infty}(n)  $ we denote a quantity
tending to infinity (for $ n \rightarrow \infty $) more slowly
than n.

 The function F is symmetric under permutations of the replicas;
 hence as long as a certain vector $ \vec{u}^{\star} $ maximizes
 F, so too does any vector $ \vec{u} $ such that $ u_{\sigma_{1} , \cdots ,
 \sigma_{r}} \; = \; u^{\star}_{\sigma_{\pi(1)}, \cdots , \sigma_{\pi(r)}
 } \; \pi \in S_{r} $.

The assumption that F has a unique maximum that is itself
invariant under replicas' permutation:
\begin{equation}
    u_{\sigma_{1} , \cdots , \sigma_{r}}^{\star} \; = \; u_{\sigma_{\pi(1)} , \cdots ,
    \sigma_{\pi(r)}}^{\star} \; \; \forall \pi \in S_{r}
\end{equation}
is known as the \emph{assumption of replica symmetry} while its
negation is called \emph{replica symmetry breaking}  since from a
group theoretical viewpoint, it   consists in a breakage of the
permutational symmetry $S_{r}$ (under which the hamiltonian $
\sum_{a=1}^{r} H( s^{a}_{1} , \cdots , s^{a}_{n})$ is of course
invariant) reducing the symmetry of the system to a suitable
subgroup $ G_{r} \subset S_{r} $.

Under the assumption of replica symmetry the computation of $
\overline{Z^{r}} $  would seem to support the \emph{satisfiability
threshold conjecture} since one finds a threshold value $
\alpha_{c} $ such that:
\begin{equation}
  \bar{E} \; = \; 0 \; \; \ for \;  \alpha < \alpha_{c}
\end{equation}
\begin{equation}
    \bar{E} \; > \; 0 \; \; for \; \alpha > \alpha_{c}
\end{equation}

For the exposed reasons, anyway, such an argument has no
mathematical consistence.

The situation is even worse as to the investigations of the phase
structure of \emph{ random k-SAT }involving \emph{replica symmetry
breaking}, a concept of which no consistent mathematical
formalization exists.

The original explanation of such a concept given by Parisi and
coworkers  in the section3.5 " Replica Symmetry Breaking: the
Final Formulation" of \cite{Mezard-Parisi-Virasoro-87} reminds
Ionesco's Absurd's Theater\footnote{so reaching a "dramatic
tension" strongly higher than the Shakespeare's one cited in the
introduction of \cite{Mezard-Parisi-Virasoro-87}.} : indeed the
breakage of the permutational symmetry $ S_{r} \rightarrow G_{r}
\subset S_{r} $ is therein augmented with a nonsense "analytic
continuation" to $ r \in \mathbb{R} $ that in the limit $ r
\rightarrow 0 $ is claimed to imply that the group of permutations
of zero objects $ S_{0} $ would contain itself as a subgroup.

Of course nothing of such an explanation is mathematical
meaningful:

if Tom has zero apples the number of ways in which he can order
them is of course zero.

Indeed, despite the many efforts to recast Parisi's theory
concerning the mean field approximation of the
Sherrington-Kirkpatrick's model into a mathematically meaningful
form \cite{Talagrand-03} (not to speak about the more critical
viewpoints concerning such a theory such as those exposed in
\cite{Newman-97}, \cite{Bolthausen-02}) the whole replica
formalism still lacks of any mathematical rigor.

In this brief notes we will present some elementary but rigorous
argument that could be useful to recast some feature of such a
formalism in a mathematically consistent framework.
\newpage
\section{Permutation group of a set}
Given a set X let us introduce the following:

\begin{definition}
\end{definition}
\emph{permutation on X:}

a bijective map $ p : X \mapsto X $

\smallskip

\begin{definition}
\end{definition}
\emph{permutation group of X:}
\begin{equation*}
   ( Perm(X) , \cdot )
\end{equation*}
where:
\begin{itemize}
    \item
\begin{equation*}
    Perm(X) \; := \; \{ p \, : \, \text{ permutation on X} \}
\end{equation*}
    \item
    $ \cdot$ is the map composition
\end{itemize}

Let us recall the following basic \cite{Isham-89}:

\begin{theorem} \label{th:Cayley theorem}
\end{theorem}
CAYLEY THEOREM:

\begin{hypothesis}
\end{hypothesis}

\begin{center}
  $G_{1}$  group
\end{center}

\begin{thesis}
\end{thesis}
\begin{equation*}
    \exists X \; set , \exists G_{2} \text{ subgroup of Perm(X)}  \; : \;
    G_{1} \sim_{is} G_{2}
\end{equation*}
where $ \sim_{is} $ denotes the isomorphism equivalence relation.

\smallskip

\begin{example}
\end{example}
Let us suppose that $ | X | = n \in \mathbb{N}_{+} $. Then $
Perm(X) = S_{n} $ is the $ n^{th}$ symmetric group. One has
clearly that $ | Perm(X) | = n ! $. Theorem \ref{th:Cayley
theorem} implies that any finite group of order n is isomorphic to
a subgroup of $ S_{n} $.

\smallskip

Let us now consider the set $ S_{0} \; := Perm(\emptyset ) $ .

One has clearly that:
\begin{equation}
    | S_{0} | \; = 0
\end{equation}
and hence:
\begin{equation}
    S_{0} \; = \; \emptyset
\end{equation}

Let us now suppose to have a set X such that $Perm(X)$ possesses
the property that Parisi and coworkers erroneously ascribe to $
S_{0} $: the property of being isomorphic to a subgroup G of
its\footnote{Actually we don't know if such a set X exists; here
we assume the existence of such a set to derive some property
that, if it exists, X must possess.}:
\begin{equation}
    Perm(X) \; \sim_{is} \; G \subset Perm(X)
\end{equation}
Since G is in particular a subset of Perm(X), Perm(X)  is
bijective to a proper subset of its and hence
\cite{Kolmogorov-Fomin-70} it is an infinite set:
\begin{equation}
  | Perm(X) | \; \geq \; \aleph_{0}
\end{equation}
from which it follows that:
\begin{equation}
    | X | \; \geq \; \aleph_{0}
\end{equation}

In particular $ X \neq S_{0} $.
\newpage
\section{One parameter families of permutation groups}

Let us consider a one-parameter family of sets $ \{ X_{\alpha} ,
\alpha \in \mathbb{R} \} $ such that:
\begin{equation}
    | X_{n} | \; = \; n \; \; \forall n \in \mathbb{N}
\end{equation}
So in particular:
\begin{equation}
    | X_{0} | \; = \; 0
\end{equation}
and hence:
\begin{equation}
  X_{0} \; = \; \emptyset
\end{equation}
Let us now observe that:
\begin{equation} \label{eq:cardinality of the permutation group for n integer}
    | Perm ( X_{n} ) | \; = \; n ! \;= \; \Gamma (n+1) \; \; \forall n \in \mathbb{N}_{+}
\end{equation}
 where:
\begin{equation}
    \Gamma (z) \; := \; \int_{0}^{\infty} t^{z-1} \exp(-t) dt \;
    \; for \; Re(z) > 0
\end{equation}
is the Euler Gamma function \cite{Hassani-99}.

Let us remark that:
\begin{equation}
   | Perm ( X_{0} ) | \; \neq \; 0 ! \;= \; \Gamma(1) \; = \; 1
\end{equation}

Clearly the right-hand side of equation\ref{eq:cardinality of the
permutation group for n integer} is well defined on the whole
interval $ ( -1 , + \infty ) $ and in particular:
\begin{equation}
   \lim_{n \rightarrow 0} \Gamma (n+1) \; = \; \Gamma (1) \; = \;
   0! \; = \; 1
\end{equation}

Let us observe anyway that:
\begin{equation}
   | Perm ( X_{\alpha} ) | \; \neq \; \Gamma (\alpha+1) \; \;
   \forall \alpha \in \mathbb{R} - \mathbb{N}
\end{equation}
since for every set S:
\begin{equation}
    | S | \; \in \; \mathbb{N} \cup \{ \aleph_{n} , n \in
    \mathbb{N} \}
\end{equation}
\newpage
\section{Consecutive replica symmetry breakings}

 Let us consider a system $ S := \{ s_{1} , \cdots ,  s_{n} \} $
consisting
 of $ n \in \mathbb{N} $  sub-systems.

 Let us suppose that initially the n sub-systems are identical.
This means that for every property P one has that:
\begin{equation}
    P( \{ s_{1} , \cdots , s_{n} \}) \; = \; P ( \{ s_{\pi(1)} , \cdots ,
    s_{\pi(n) } \} )
 \; \; \forall \pi \in S_{n}
\end{equation}
In physical terms this means that $ S_{n} $ is a symmetry of the
system.

 Given a number $ m_{1} \in \mathbb{N} \; : \; \frac{n}{m_{1}}
\in \mathbb{N} $ let us divide the system S in $ \frac{n}{ m_{1}}
$ groups $ g^{1}_{1} := \{ s_{1} , \cdots , s_{m_{1}} \} , \cdots
, g^{1}_{\frac{n}{m_{1}}} := \{ s_{n-m_{1}} , \cdots , s_{n} \} $
each consisting of $ m_{1} $ elements.

Let us now suppose to differentiate the systems belonging to a
group $ g^{1}_{i} $ from those belonging to a  different group $
g^{1}_{j} \, i \neq j $; this means that for every property P:
\begin{equation}
    P ( x ) = P( y ) \; \; \forall x , y \in
    g^{1}_{i} , \forall i= 1, \cdots , \frac{n}{m_{1}}
\end{equation}
but that exists a property P such that:
\begin{equation}
   P ( x ) \neq P( y ) \; \; \forall x \in g^{1}_{i}
   , y \in  g^{1}_{j} \; : \; i \neq j
\end{equation}

 In physical terms this means to perform the symmetry breaking $
S_{n} \mapsto  G_{n;m_{1}}  $ where:
\begin{equation}
   G_{n;m_{1}} \; := \; ( S_{m_{1}})^{\frac{n}{m_{1}}
}
\end{equation}

In fact the system is now symmetric only under the  $ m_{1} !$
permutations of the elements inside each group.

Clearly:
\begin{equation}
    | G_{n;m_{1}} | \; = \; ( m_{1}!)^{\frac{n}{m_{1}}}
\end{equation}

Let us observe that, contrary to what is claimed in the section
3.5 "Replica Symmetry Breaking: the Final Formulation" of
\cite{Mezard-Parisi-Virasoro-87}, the system now is not invariant
under the $ (\frac{n}{m_{1}})!$ permutations of the groups $
g^{1}_{1}, \cdots , g^{1}_{\frac{n}{m_{1}}} $  since now there
exists a property distinguishing these groups.

\smallskip

Given a number $ m_{2} \in \mathbb{N} $ such that $
\frac{m_{1}}{m_{2}} \in \mathbb{N} $  let us divide each group $
g^{1}_{i} $ in $ \frac{m_{1}}{m_{2}}$ sub-groups each consisting
of $ m_{2} $ elements; so the group $ g^{1}_{1} $ is divided in
the subgroups $ g^{2}_{1} , \cdots , g^{2}_{\frac{m_{1}}{m_{2}}} $
and so on.

Let us now suppose to differentiate the systems belonging to a
group $ g^{2}_{i} $ from those belonging to a  different group $
g^{2}_{j} \, i \neq j $; this means that for every property P:
\begin{equation}
    P ( x ) = P( y ) \; \; \forall x , y \in
    g^{2}_{i} , \forall i= 1, \cdots , \frac{n}{ m_{2}}
\end{equation}
but that exists a property P such that:
\begin{equation}
   P ( x ) \neq P( y ) \; \; \forall s_{1} \in g^{2}_{i}
   , s_{2} \in  g^{2}_{j} \; : \; i \neq j
\end{equation}

In physical terms this means to perform the symmetry breaking $
G_{n;m_{1}} \mapsto   G_{n;m_{2}} $ where of course:
\begin{equation}
    G_{n;m_{2}} \; := \; (
    S_{m_{2}})^{\frac{n}{m_{2}}}
\end{equation}

Such a procedure can be iterated a certain number of times;
supposed that there exist  $ k+1 \in \mathbb{N} $ natural numbers
$ m_{1} , \cdots , m_{k} $ such that:
\begin{equation}
    \frac{n}{m_{1}} \in \mathbb{N} \; and \; \frac{m_{i}}{m_{i+1}} \in \mathbb{N} \; \; i=1, \cdots , k
\end{equation}
at the $ k^{th}$ step one performs the symmetry breaking:
\begin{equation}
    G_{n;m_{k}} \rightarrow G_{n;m_{k+1}}
\end{equation}
where of course:
\begin{equation}
  G_{n;m_{k}} \; := \; (
    S_{m_{k}})^{\frac{n}{m_{k}}}
\end{equation}
\begin{equation}
  G_{n;m_{k+1}} \; := \; (
    S_{m_{k+1}})^{\frac{n}{m_{k+1}}}
\end{equation}

Let us observe, anyway, that exists a maximum number $ k_{max} \in
\mathbb{N} $ of possible consecutive replica symmetry breakings.

If one was interested in performing the limit $ n \rightarrow
\infty $ one could argue that the maximum number of possible
consecutive replica symmetry breaking "tends to infinity" in the
following sense: for every $ k \in \mathbb{N} $ there exists an $
n \in \mathbb{N} $ and k+1 numbers $ m_{1}, \cdots , m_{k+1} \in
\mathbb{N} $ such that:
\begin{equation}
    \frac{n}{m_{1}} \in \mathbb{N} \; and \; (  \frac{m_{i}}{m_{i+1}} \in \mathbb{N} \; \; i=1, \cdots ,
    k)
\end{equation}

Unfortunately Parisi's theory involves instead a mathematically
inconsistent limit of the iteration's procedure for $ n
\rightarrow 0$.

\end{document}